\documentclass[runningheads]{llncs}

\usepackage{graphicx}
\usepackage{algorithm}
\usepackage{algorithmic}
\usepackage{amsmath,amssymb}
\usepackage{mathrsfs}
\usepackage[section]{placeins}
\usepackage{subfigure}
\usepackage{booktabs}
\usepackage{multirow}
\usepackage{float}
\usepackage[dvipsnames]{xcolor}
\usepackage{wrapfig}

\begin{document}

\title{CDiffMR: Can We Replace the Gaussian Noise with K-Space Undersampling for Fast MRI?}

\titlerunning{CDiffMR: Cold Diffusion for MRI Recon}

\author{Jiahao Huang\inst{1, 2}\and
Angelica I. Aviles-Rivero\inst{3} \and
Carola-Bibiane Sch{\"o}nlieb\inst{3} \and
Guang Yang\inst{1, 2}}

\authorrunning{J. Huang et al.}

\institute{
National Heart and Lung Institute, Imperial College London, London, United Kingdom \\
\email{\{j.huang21, g.yang\}@imperial.ac.uk} 
\and Cardiovascular Research Centre, Royal Brompton Hospital, London, United Kingdom 
\and Department of Applied Mathematics and Theoretical Physics, University of Cambridge, United Kingdom}

\maketitle              
\begin{abstract}

Deep learning has shown the capability to substantially accelerate MRI reconstruction while acquiring fewer measurements. 
Recently, diffusion models have gained burgeoning interests as a novel group
of deep learning-based generative methods. These methods seek to sample data points that belong to a target distribution from a Gaussian distribution, which has been successfully extended to MRI reconstruction.
In this work, we proposed a Cold Diffusion-based MRI reconstruction method called CDiffMR.
Different from conventional diffusion models, the degradation operation of our CDiffMR is based on \textit{k}-space undersampling instead of adding Gaussian noise, and the restoration network is trained to harness a de-aliaseing function. 
We also design starting point and data consistency conditioning strategies to guide and accelerate the reverse process.
More intriguingly, the pre-trained CDiffMR model can be reused for reconstruction tasks with different undersampling rates.
We demonstrated, through extensive numerical and visual experiments, that the proposed CDiffMR can achieve comparable or even superior reconstruction results than state-of-the-art models. Compared to the diffusion model-based counterpart, CDiffMR reaches readily competing results using only $1.6 \sim 3.4\%$ for inference time. The code is publicly available at https://github.com/ayanglab/CDiffMR. 

\end{abstract}

\section{Introduction}

Magnetic Resonance Imaging (MRI) is an essential non-invasive technique that enables high-resolution and reproducible assessments of structural and functional information, for clinical diagnosis and prognosis, without exposing the patient to radiation.
Despite its widely use in clinical practice, MRI still suffers from the intrinsically slow data acquisition process, which leads to uncomfortable patient experience and artefacts from voluntary and involuntary physiological movements~\cite{Chen_2022_AI}.

\begin{figure}[t]
    \centering
    \includegraphics[width=5in]{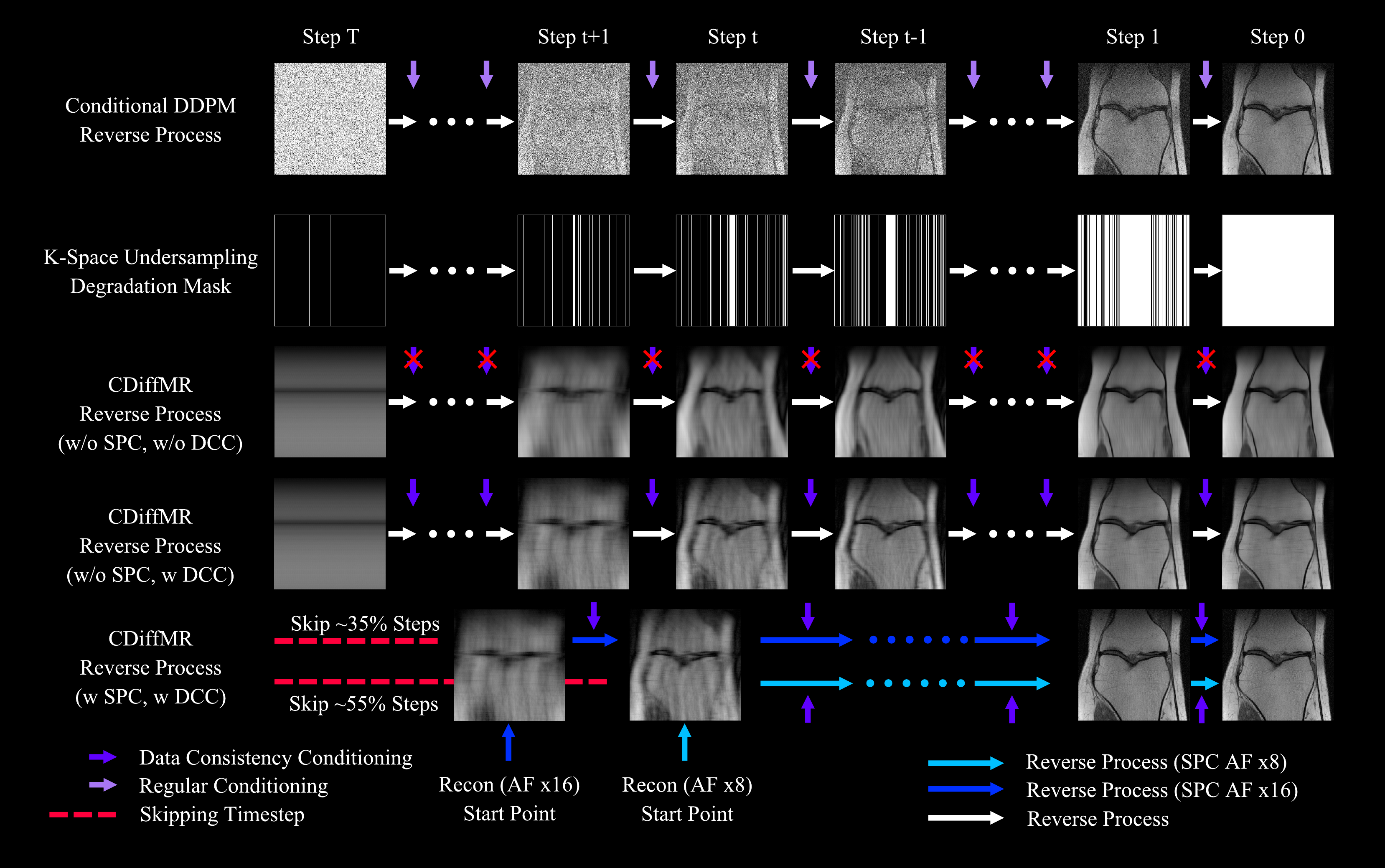}
    \caption{
    Row 1: The reverse process of conditional DDPM~\cite{Ho_2022_DDPM};
    Row 2: The K-Space Undersampling Degradation Mask for the proposed CDiffMR;
    Row 3: The reverse process of CDiffMR without DDC and SPC;
    Row 4: The reverse process of CDiffMR with DDC but without SPC;
    Row 5: The reverse process of CDiffMR with DDC and SPC;
    SPC: Starting Point Conditioning;
    DDC: Data Consistency Conditioning.
    }
    \label{fig:FIG_INTRO}
\end{figure}

Deep learning has achieved considerable success across various research domains in recent years, including the ability to substantially accelerate MRI reconstruction while requiring fewer measurements.
Various kinds of deep learning-based models, including Convolutional Neural Networks~\cite{Schlemper_2017_D5C5,Yang_2018_DAGAN}, Recurrent Neural Networks~\cite{Chen_2022_Pyramid,Guo_2021_Over}, Graph Neural Networks~\cite{Huang_2023_VIGU} or Transformers~\cite{Huang_2022_SwinMR,Korkmaz_2022_Unsupervised}, have been explored for MRI reconstruction and achieved impressive success with a high accelerate factor (AF).
However, most of these methods are based on a strong degradation prior, i.e., the undersampling mask, which entails a performance drop when the training and testing undersampling masks mismatch~\cite{Peng_2022_Towards,Efrat_2022_Implicit}. Therefore, additional training is required when applying different undersampling mask condition, leading to a waste of computational resources. 

Diffusion models~\cite{Ho_2022_DDPM,Song_2019_Score,Song_2020_SDE} represent a group of unconditional generative methods which sample data points that belong to target distribution from a Gaussian distribution. The earliest diffusion models type was known as Denoising Diffusion Probabilistic Models (DDPMs)~\cite{Ho_2022_DDPM} and Score Matching with Langevin Dynamics (SMLD)~\cite{Song_2019_Score}, which were later unified into a framework by Score-based Stochastic Differential Equation (SDE)~\cite{Song_2020_SDE}. 

Diffusion models have been widely applied for inverse problems~\cite{Song_2021_Solving,Chung_2022_CCDF} including MRI Reconstruction~\cite{Peng_2022_Towards,Chung_2022_Score,Cao_2022_Accelerating,Gungor_2022_Adaptive,Cao_2022_High}.
Peng et al.~\cite{Peng_2022_Towards} proposed a diffusion model-based MR reconstruction method, called DiffuseRecon, which did not require additional training on specific acceleration factors. 
Chung et al.~\cite{Chung_2022_Score} designed a score-based model for MRI reconstruction, which performed the reconstruction task iteratively using a numerical SDE solver and data consistency step. 
Cao et al.~\cite{Cao_2022_Accelerating} proposed a complex diffusion probabilistic model for MRI reconstruction for better preservation of the MRI complex-value information. 
Gungor et al.~\cite{Gungor_2022_Adaptive} introduced an adaptive diffusion prior, namely AdaDiff, for enhancing reconstruction performance during the inference stage.
Cao et al.~\cite{Cao_2022_High} designed a modified high-frequency DDPM model for high-frequency information preservation of MRI data.
However, they do share a commonality -- the prolonged inference time due to the iterative nature of diffusion models. Chung et al.~\cite{Chung_2022_CCDF} proposed a new reverse process strategy for accelerating the sampling for the reverse problem, Come-Closer-Diffuse-Faster (CCDF), suggesting that \textit{starting from Gaussian noise is necessary for diffusion models}. CCDF-MRI achieved outstanding reconstruction results with reduced reverse process steps. 

Most existing diffusion models, including the original DDPM, SMLD and their variants, are strongly based on the use of Gaussian noise, which provides the `random walk' for `hot' diffusion. 
Cold Diffusion Model~\cite{Bansal_2022_Cold} rethought the role of the Gaussian noise, and generalised the diffusion models using different kinds of degradation strategies, e.g., blur, pixelate, mask-out, rather than the Gaussian noise applied on conventional diffusion models.

In this work, a novel Cold Diffusion-based MRI Reconstruction method (CDiffMR) is proposed (see Fig.~\ref{fig:FIG_INTRO}). 
CDiffMR introduces a novel K-Space Undersampling Degradation (KSUD) module for the degradation, which means CDiffMR does not depend on the Gaussian noise. Instead of building an implicit transform to target distribution by Gaussian noise, CDiffMR explicitly learns the relationship between undersampled distribution and target distribution by KSUD.

We propose two novel \textit{k}-space conditioning strategies to guide the reverse process and to reduce the required time steps.
1) Starting Point Conditioning (SPC). The \textit{k}-space undersampled zero-filled images, which is usually regarded as the network input, can act as the reverse process starting point for conditioning. The number of reverse time steps therefore depends on the undersamping rate, i.e., the higher \textit{k}-space undersampling rate (lower AF, easier task), the fewer reverse time steps required.
2) Data Consistency Conditioning (DCC). In every step of the reverse process, data consistency is applied to further guide the reverse process in the correct way.

It is note that our CDiffMR is a one-for-all model. This means that once CDiffMR is trained, it can be reused for all the reconstruction tasks, with any reasonable undersampling rates conditioning, as long as the undersampling rate is larger than the preset degraded images $x_T$ at the end of forward process (e.g., $1\%$ undersampling rate).
Experiments were conducted on FastMRI dataset~\cite{Zbontar_2018_fastMRI}. The proposed CDiffMR achieves comparable or superior reconstruction results with respect to state-of-the-art methods, and reaches a much faster reverse process compared with diffusion model-based counterparts.
For the sake of clarity, we use `sampling' or `undersampling' to specify the \textit{k}-space data acquisition for MRI, and use `reverse process' to represent sampling from target data distribution in the inference stage of diffusion models.
Our main contributions are summarised as follows:
\begin{itemize}
\item[$\bullet$] 
An innovative Cold Diffusion-based MRI Reconstruction methods is proposed. To best of our knowledge, CDiffMR is the first diffusion model-based MRI reconstruction method that exploits the \textit{k}-space undersampling degradation.
\item[$\bullet$]
Two novel \textit{k}-space conditioning strategies, namely SPC and DCC, are developed to guide and accelerate the reverse process.
\item[$\bullet$]
The pre-trained CDiffMR model can be reused for MRI reconstruction tasks with a reasonable range of undersampling rates.
\end{itemize}

\section{Methodology}

This section details two key parts of the proposed CDiffMR: 1) the optimisation and training schemes and 2) the \textit{k}-space conditioning reverse process.

\subsection{Model Components and Training}

Diffusion models are generally composed of a degradation operator $\operatorname{D}(\cdot, t)$ and a learnable restoration operator $\operatorname{R}_\theta(\cdot, t)$~\cite{Bansal_2022_Cold}. 
For standard diffusion models, the $\operatorname{D}(\cdot, t)$ disturbs the images via Gaussian noise according to a preset noise schedule controlled by time step $t$. The $\operatorname{R}_\theta(\cdot, t)$ is a denoising function controlled by $t$ for various noise levels. 

\begin{wrapfigure}{R}{0.5\textwidth}
    \centering
    \vspace{-0.8cm}
    \includegraphics[width=0.45\textwidth]{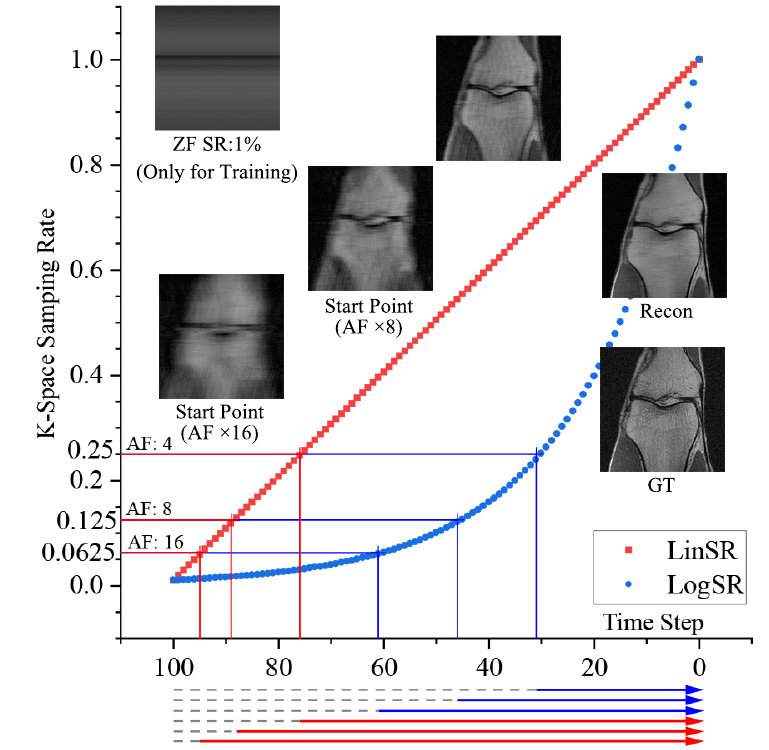}
    \caption{Linear and Log \textit{k}-space undersampling degradation schedules of CDiffMR.}
    \vspace{-0.3cm} 
    \label{fig:FIG_Schedule}
\end{wrapfigure}

Instead of applying Gaussian noise, CDiffMR provides a new option, namely KSUD operator $\operatorname{D}(\cdot, t)$ and de-aliasing restoration operator $\operatorname{R}_\theta(\cdot, t)$. 
With the support of KSUD, CDiffMR can explicitly learn the relationship between input distribution of \textit{k}-space undersampled images and target distribution of fully-sampled images that is built implicitly by Gaussian noise in the conventional diffusion model. 
The KSUD operator $\operatorname{D}(x, t) = \mathcal{F}^{-1} \mathcal{M}_{t} \mathcal{F} x$, undersamples input images via different \textit{k}-space mask of which undersampling rate is controlled by the time step $t$.

Two \textit{k}-space sampling rate $\mathrm{SR}_{t}$ schedule are designed in this study, linear (LinSR) and log (LogSR) sampling rate schedules. We set $\mathrm{SR}_{t}=1 - (1 - \mathrm{SR}_\mathrm{min})\frac{t}{T}$ for LinSR schedule, and $\mathrm{SR}_{t} = \mathrm{SR}_\mathrm{min}^{\frac{t}{T}}$ for LogSR schedule (see Fig.~\ref{fig:FIG_Schedule}). $\operatorname{D}(x, t+1)$ contains less \textit{k}-space information compared to $\operatorname{D}(x, t)$, and when $t=0$ we have:
\begin{equation}\label{eq:degradation_t_0}
\begin{aligned}
x = \operatorname{D}(x, 0) &= \mathcal{F}^{-1} \mathcal{M}_{0} \mathcal{F} x = \mathcal{F}^{-1} \mathcal{I} \mathcal{F} x,
\end{aligned}
\end{equation}
\noindent where $\mathcal{M}_{t}$ is the undersampling mask at step $t$ corresponding to $\mathrm{SR}_{t}$, and $\mathcal{M}_{0} = \mathcal{I}$ is the fully-sampling mask (identity map). $\mathcal{F}$ and $\mathcal{F}^{-1}$ denote Fourier and inverse Fourier transform.

The restoration operator $\operatorname{R}_\theta(\cdot, t)$ is an improved U-Net with time embedding module, following the official implementation\footnote{https://github.com/ermongroup/ddim} of Denoising Diffusion Implicit Models~\cite{Song_2020_DDIM}. An ideal $\operatorname{R}_\theta(\cdot, t)$ should satisfy $x_0 \approx \operatorname{R}_\theta(x_t, t)$.

For the training of the restoration operator $\operatorname{R}_\theta(\cdot, t)$, $x_{\mathrm{true}}$ is the fully-sampled images randomly sampled from target distribution $\mathcal{X}$. Practically, time step $t$ is randomly chosen from $(1, T]$ during the training. The driven optimisation scheme reads:
\begin{equation}\label{eq:training}
\begin{aligned}
\min_\theta \mathbb{E}_{x_{\mathrm{true}} \sim \mathcal{X}}
\|\operatorname{R}_\theta(\operatorname{D}(x_{\mathrm{true}}, t), t)-x\|, 
\quad\quad t = 1 ... T.
\end{aligned}
\end{equation}

\subsection{K-Space Conditioning Reverse Process}

Two \textit{k}-space conditioning strategies, SPC and DCC, are designed to guide and accelerate the reverse process.

Starting Point Conditioning enables the reverse process of CDiffMR to start from the \emph{half way} step $T^\prime$ instead of step $T$ (see Fig.~\ref{fig:FIG_INTRO} and Fig.~\ref{fig:FIG_Schedule}).
The starting point of the reverse process depends on the \textit{k}-space undersampling rate of the reconstruction task. 
Specifically, for the reconstruction task with $\mathcal{M}$, $T^{\prime}$ can be checked by comparing the sampling rate of $\mathcal{M}$ in the degradation schedule, and the corresponding reverse process start step $T^{\prime}$ can be located, which is expressed as:
\begin{equation}\label{eq:locate_start_point}
\begin{aligned}
\text{Sampling Rate:}\quad \mathcal{M}_{T} < \mathcal{M}_{T^{\prime}} < \mathcal{M} < \mathcal{M}_{T^{\prime} - 1} < \mathcal{M}_{0} = \mathcal{I}.
\end{aligned}
\end{equation}

With the start step $T^{\prime}$, the reverse process is conditioned by the reverse process of the initial image $x_{T^{\prime}} \gets \mathcal{F}^{-1} y$. The framework of the reverse process follows Algorithm 2 in~\cite{Bansal_2022_Cold}, whereas we applied DCC strategy for further guiding (Eq.~\eqref{eq:reverse_2}). 
The result of the reverse process $x_{0}$ is the final reconstruction result. The whole reverse process is formulated as:
\begin{equation}\label{eq:reverse_1}
\begin{aligned}
\hat x_{0,t}^{\prime} &= \operatorname{R}_\theta(x_{t}, t), \quad \text{s.t. } t = T^{\prime} ... 1.
\end{aligned}
\end{equation}
\begin{equation}\label{eq:reverse_2}
\begin{aligned}
\hat x_{0,t} &= \mathcal{F}^{-1} (1 - \mathcal{M}) \mathcal{F} \hat x_{0,t}^{\prime} + \mathcal{F}^{-1} \mathcal{M} \mathcal{F} x_{T^{\prime}}, \quad \text{s.t. } t = T^{\prime} ... 1.
\end{aligned}
\end{equation}
\begin{equation}\label{eq:reverse_3}
\begin{aligned}
x_{t-1} &= x_{t} - \operatorname{D}(\hat x_{0,t} , t) + \operatorname{D}(\hat x_{0,t} , t-1), \quad \text{s.t. } t = T^{\prime} ... 1.
\end{aligned}
\end{equation}

\section{Experimental Results}

This section describes in detail the set of experiments conducted to validate the proposed CDiffMR.

\begin{figure}[t]
    \centering
    \includegraphics[width=4.5in]{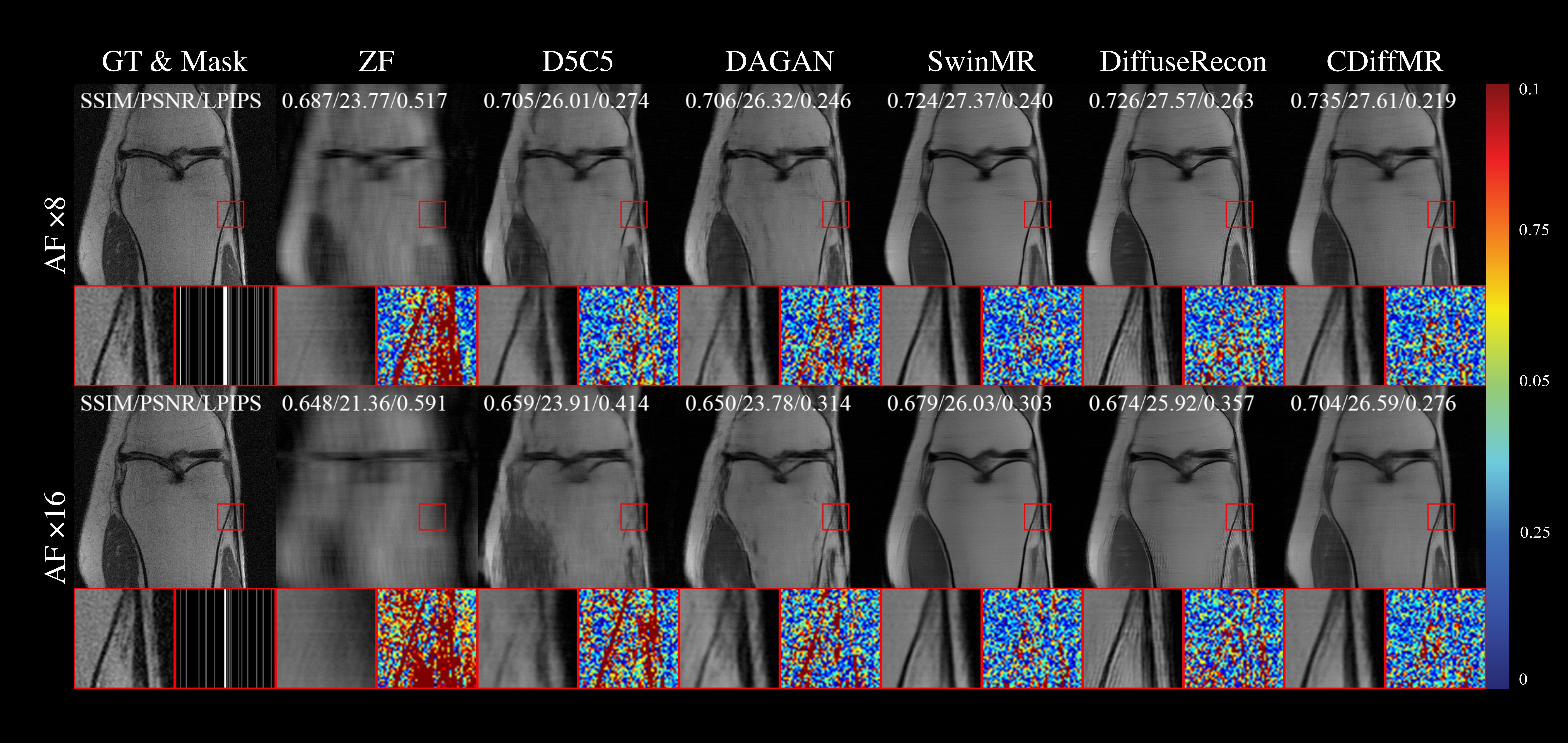}
    \caption{
    Visual comparison against the state-of-the-art approaches with accelerate rate (AF) $\times 8$ and $\times 16$. The colour bar denotes the difference between the reconstructed result and ground truth. Zoom-in views displays selected fine detailed regions.
    }
    \label{fig:FIG_MainExperiment_Row2}
\end{figure}

\subsection{Implementation Details and Evaluation Methods}


The experiments were conducted on the FastMRI dataset~\cite{Zbontar_2018_fastMRI}, which contains single-channel complex-value MRI data.
For the FastMRI dataset, we applied 684 proton-density weighted knee MRI scans without fat suppression from the official training and validation sets, which were randomly divided into training set (420 cases), validation set (64 cases) and testing set (200 cases), approximately according to a ratio of 6:1:3. 
For each case, 20 coronal 2D single-channel complex-value slices near the centre were chosen, and all slices were centre-cropped to $320 \times 320$. 

Undersampling mask $\mathcal{M}$ and $\mathcal{M}_{t}$ were generated by the fastMRI official implementation. We applied AF$\times 8$ and $\times 16$ Cartesian mask for all experiments.


Our proposed CDiffMR was trained on two NVIDIA A100 (80GB) GPUs, and tested on an NVIDIA RTX3090 (24GB) GPU. CDiffMR was trained for 100,000 gradient steps, using the Adam optimiser with a learning rate 2e-5 and a batch size 24.
We set the total diffusion time step to $T=100$ for both LinSR and LogSR degradation schedules. The minimal sampling rate (when $t=100$) was set to $1\%$.

For comparison, we used CNN-based methods D5C5~\cite{Schlemper_2017_D5C5}, DAGAN~\cite{Yang_2018_DAGAN}, Transformers-based method SwinMR~\cite{Huang_2022_SwinMR}, novel diffusion model-based method DiffuseRecon~\cite{Peng_2022_Towards}.
We trained D5C5 and DiffuseRecon following the official setting, while we modified DAGAN and SwinMR for 2-channel input, output and loss function, as they were officially proposed for single-channel reconstruction. 

In the quantitative experiments, Peak Signal-to-Noise Ratio (PSNR), Structural Similarity Index Measure (SSIM), Learned Perceptual Image Patch Similarity (LPIPS)~\cite{Zhang_2018_LPIPS} were applied to examine the reconstruction quality. Among them, LPIPS is a deep learning-based perceptual metric, which can match human visual perception well.
The inference time was measured on a NVIDIA RTX3090 (24GB) GPU with an input shape of $(1, 1, 320, 320)$ for ten times.

\subsection{Comparison and Ablation Studies}

The quantitative results are reported in Table~\ref{tab:Quantitative_Result} and further supported by visual comparisons in Fig.~\ref{fig:FIG_MainExperiment_Row2}.
The proposed CDiffMR achieves promising results compared to the SOTA MRI reconstruction methods.
Compared with the diffusion model-based method DiffuseRecon, CDiffMR achieves comparable or better results with only $1.6 \sim 3.4\%$ inference time of DiffuseRecon. 
For ablation studies, we explored how DDC and SPC affect the speed of reverse process and reconstruction quality (see Fig.~\ref{fig:FIG_Ablation}(A)(B)).

\begin{table}[!t]
  \centering
  \caption{
  Quantitative reconstruction results with accelerate rate (AF) $\times 8$ and $\times 16$. LinSR (LogSR): linear (log) sampling rate schedule;
  $^{\star}$($^{\dagger}$): significantly different from CDiffMR-LinSR(LogSR) by Mann-Whitney Test.
  }
  \scalebox{0.8}{
    \setlength{\tabcolsep}{4mm}{
    \begin{tabular}{ccccc}
    \toprule
    Model (AF $\times 8$) & SSIM $\uparrow$  & PSNR $\uparrow$  & LPIPS $\downarrow$ & Inference Time $\downarrow$ \\
    \midrule
    ZF    & 0.678 (0.087)$^{\star}$$^{\dagger}$ & 22.74 (1.73)$^{\star}$$^{\dagger}$ & 0.504 (0.058)$^{\star}$$^{\dagger}$ & -- \\
    D5C5~\cite{Schlemper_2017_D5C5}  & 0.719 (0.104)$^{\star}$$^{\dagger}$ & 25.99 (2.13)$^{\star}$$^{\dagger}$ & 0.291 (0.039)$^{\star}$$^{\dagger}$ & 0.044 \\
    DAGAN~\cite{Yang_2018_DAGAN} & 0.709 (0.095)$^{\star}$$^{\dagger}$ & 25.19 (2.21)$^{\star}$$^{\dagger}$ & 0.262 (0.043)$^{\star}$$^{\dagger}$ & 0.004 \\
    SwinMR~\cite{Huang_2022_SwinMR} & 0.730 (0.107)$^{\star}$$^{\dagger}$ & 26.98 (2.46)$^{\star}$$^{\dagger}$ & 0.254 (0.042)$^{\star}$$^{\dagger}$ & 0.037 \\
    DiffuseRecon~\cite{Peng_2022_Towards} & 0.738 (0.108)$^{\star}$ & \textbf{27.40 (2.40)} & 0.286 (0.038)$^{\star}$$^{\dagger}$ & 183.770 \\
    \midrule
    CDiffMR-LinSR & \textbf{0.745 (0.108)} & \underline{27.35 (2.56)} & \underline{0.240 (0.042)} & 5.862 \\
    CDiffMR-LogSR & \underline{0.744 (0.107)} & 27.26 (2.52) & \textbf{0.236 (0.041)} & 3.030 \\
    \bottomrule
    Model (AF $\times 16$) & SSIM $\uparrow$  & PSNR $\uparrow$  & LPIPS $\downarrow$ & Inference Time $\downarrow$ \\
    \midrule
    ZF    & 0.624 (0.080)$^{\star}$$^{\dagger}$ & 20.04 (1.59)$^{\star}$$^{\dagger}$ & 0.580 (0.049)$^{\star}$$^{\dagger}$ & -- \\
    D5C5~\cite{Schlemper_2017_D5C5}  & 0.667 (0.108)$^{\star}$$^{\dagger}$ & 23.35 (1.78)$^{\star}$$^{\dagger}$ & 0.412 (0.049)$^{\star}$$^{\dagger}$ & 0.044 \\
    DAGAN~\cite{Yang_2018_DAGAN} & 0.673 (0.102)$^{\star}$$^{\dagger}$ & 23.87 (1.84)$^{\star}$$^{\dagger}$ & 0.317 (0.044)$^{\star}$$^{\dagger}$ & 0.004 \\
    SwinMR~\cite{Huang_2022_SwinMR} & 0.673 (0.115)$^{\star}$$^{\dagger}$ & 24.85 (2.12)$^{\star}$$^{\dagger}$ & 0.327 (0.045)$^{\star}$$^{\dagger}$ & 0.037 \\
    DiffuseRecon~\cite{Peng_2022_Towards} & 0.688 (0.119)$^{\star}$$^{\dagger}$ & 25.75 (2.15)$^{\star}$$^{\dagger}$ & 0.362 (0.047)$^{\star}$$^{\dagger}$ & 183.770 \\
    \midrule
    CDiffMR-LinSR & \textbf{0.709 (0.117)} & \textbf{25.83 (2.27)} & \underline{0.297 (0.042)} & 6.263 \\
    CDiffMR-LogSR & \underline{0.707 (0.116)} & \underline{25.77 (2.25)} & \textbf{0.293 (0.042)} & 4.017 \\
    \bottomrule
    \end{tabular}%
    }}%
  \label{tab:Quantitative_Result}%
\end{table}%

\section{Discussion and Conclusion}

\begin{figure}[ht]
    \centering
    \includegraphics[width=5in]{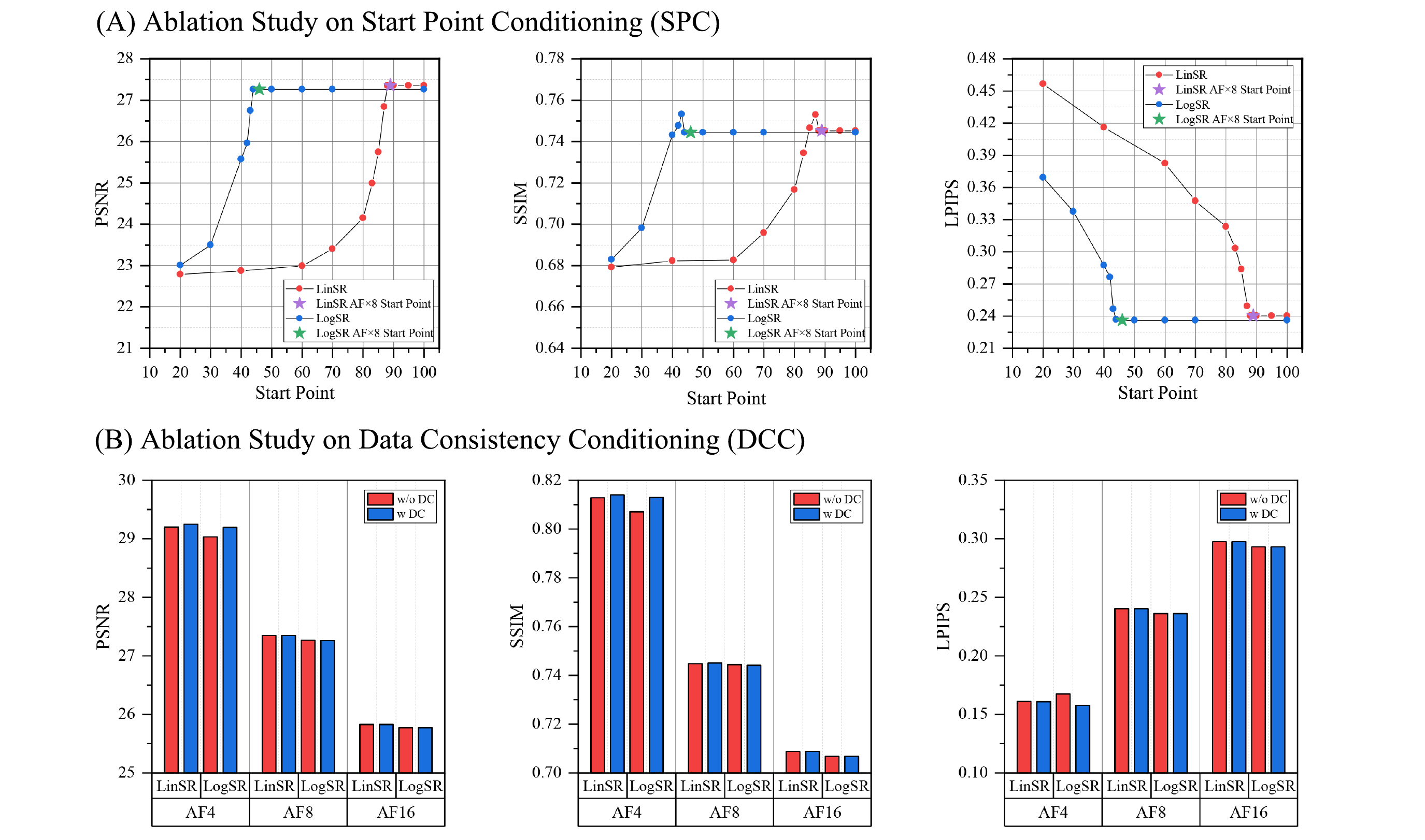}
    \caption{
    (A) The relationship between reconstruction metrics and the reverse process starting point conditioning (SPC); 
    (B) The relationship between reconstruction metrics and the reverse process data consistency conditioning (DDC).
    }
    \vspace{-0.25cm}
    \label{fig:FIG_Ablation}
\end{figure}

This work has exploited Cold Diffusion-based model for MRI reconstruction and proposed CDiffMR. 
We have designed the novel KSUD for degradation operator $\operatorname{D}(\cdot, t)$ and trained a restoration function $\operatorname{R}_\theta(\cdot, t)$ for de-aliasing under various undersampling rates. 
We pioneered the harmonisation of the degradation function in reverse problem (\textit{k}-space undersampling in MRI reconstruction) and the degradation operator in diffusion model (KSUD). In doing so, CDiffMR is able to explicitly learn the \textit{k}-space undersampling operation to further improve the reconstruction, while providing the basis for the reverse process acceleration. Two \textit{k}-space conditioning strategies, SPC and DCC, have been designed to guide and accelerate the reverse process. 
Experiments have demonstrated that \textit{k}-space undersampling can be successfully used as degradation in diffusion models for MRI reconstruction. 

In this study, two KSUD schedules, i.e., have been designed for controlling the \textit{k}-space sampling rate of every reverse time steps.
According to Table~\ref{tab:Quantitative_Result}, LogSR schedule achieves better perceptual score while LinSR has better fidelity score, where the difference is actually not significant. 
However, the required reverse process steps of LogSR is much fewer than LinSR's, which significantly accelerates the reverse process. 
This is because for LogSR schedule, a larger proportion steps are corresponding to lower sampling rate (high AF), therefore the starting point of LogSR is closer to step 0 than LinSR (see Fig.~\ref{fig:FIG_Schedule} and Fig.~\ref{fig:FIG_Ablation} (A)).
For AF$\times 16$ reconstruction task, CDiffMR-LinSR theoretically requires 95 reverse steps, while CDiffMR-LogSR only requires 61 reverse steps, and for AF$\times 8$ reconstruction task, CDiffMR-LinSR requires 89 reverse steps, while CDiffMR-LogSR only requires 46 reverse steps. 
The lower AF of the reconstruction task, the less reverse process steps required. 
Therefore, we recommend CDiffMR-LogSR as it achieves similiar results of CDiffMR-LinSR with much faster reverse process.

In the ablation studies, we have further examined the selection of reverse process starting point $T^{\prime}$. 
Fig.~\ref{fig:FIG_Ablation} (A) has shown the reconstruction quality using different starting point.
Reconstruction performance keeps stable with a range of reverse process starting points, but suddenly drops after a tuning point, which exactly matches the theoretical starting point. This experiment has proven the validity of our starting point selection method (Eq.~\eqref{eq:locate_start_point}), and shown that our theoretical starting point keeps optimal balance between the reconstruction process and reverse process speed.

We also explored the validity of DDC in the ablation studies. 
Fig.~\ref{fig:FIG_Ablation} (B) has shown the reconstruction quality with or without DDC using different sampling rate schedule with different AF. 
The improvement by the DDC is significant with a lower AF ($\times 4$), but limited with a higher AF ($\times8$, $\times16$). Therefore, CDiffMR keeps the DDC due to its insignificant computational cost.

The proposed CDiffMR heralds a new kind of diffusion models for solving inverse problems, i.e., applying the degradation model in reverse problem as the degradation module in diffusion model. 
CDiffMR has proven that this idea performs well for MRI reconstruction tasks. We can envision that our CDiffMR can serve as the basis for general inverse problems.

%

\bibliographystyle{splncs04}
\bibliography{references.bib}
\end{document}


\title{Supplementary Materials for ``CDiffMR: Can We Replace the Gaussian Noise with K-Space Undersampling for Fast MRI?"}

\titlerunning{Supplementary Materials}




\author{***}

\authorrunning{*** et al.}

\institute{***}

\maketitle              